\documentstyle[12pt,aaspp4]{article}
\tightenlines

\def	\beq	{\begin{equation}}
\def	\cm	{\,{\rm cm}}
\def	\debye	{\,{\rm debye}}
\def	\eeq	{\end{equation}}
\def	\erg	{\,{\rm ergs}}

\def	\gtsim	{\gtrsim}					  
\def	\H	{{\rm H}}
\def	\HH	{{\rm H}_2}

\def	\K	{\,{\rm K}}
\def	\Lsol	{L_{\odot}}
\def	\ltsim	{\lesssim}					  
\def	\micron	{\,\mu{\rm m}}
\def	\pc	{\,{\rm pc}}
\def	\s	{\,{\rm s}}
\def	\sr	{\,{\rm sr}}

\lefthead{Draine \& Lazarian}
\righthead{Emission from Spinning Dust Grains}

\begin{document}

\title{Diffuse Galactic Emission from Spinning Dust Grains}

\author{B.T. Draine \& A. Lazarian}
\affil{Princeton University Observatory, Peyton Hall, Princeton,
NJ 08544;
draine@astro.princeton.edu, lazarian@astro.princeton.edu}

\begin{abstract}

Spinning interstellar dust grains produce detectable
rotational emission in the 10-100 GHz frequency range.
We calculate the emission spectrum, and show
that this emission can account for the 
``anomalous'' Galactic background component which 
correlates with $100\micron$ thermal emission from dust.
Implications for cosmic background studies are discussed.
\end{abstract}


\section{Introduction}

Experiments to map the cosmic background radiation (CBR) have stimulated renewed
interest in diffuse Galactic emission.
Sensitive observations of variations in the microwave sky brightness have
revealed a component of the 14-90 GHz microwave continuum
which is correlated with
$100\micron$ thermal emission from interstellar dust
(Kogut et al.\ 1996;
de Oliveira-Costa et al.\ 1997;
Leitch et al.\ 1997).
The origin of this ``anomalous" emission has been of great interest.
While the observed frequency-dependence appears consistent
with free-free emission
(Kogut et al.\ 1996; Leitch et al.\ 1997),
it is difficult to reconcile the observed intensities with free-free
emission from interstellar gas (see \S\ref{sec:freefree} below).

A recent investigation of the rotational dynamics of very small interstellar
grains (Draine \& Lazarian 1998; hereinafter DL98) 
leads us instead to propose that this 10-100 GHz
``anomalous'' component of the diffuse Galactic background 
is produced by electric
dipole rotational emission from very small dust grains under normal
interstellar conditions.

Below we describe briefly our assumptions concerning the interstellar
grain properties (\S\ref{sec:grain_props}), the dynamics governing the
rotation of small grains (\S\ref{sec:rotation}), the predicted emission
spectrum and intensity (\S\ref{sec:prediction}) and how it compares with
observations (\S\ref{sec:observations}), and why the observed emission cannot
be attributed to free-free (\S\ref{sec:freefree}).
In \S\ref{sec:discussion} we discuss implications for future CBR studies.

\section{Grain Properties\label{sec:grain_props}}

The observed emission from interstellar diffuse clouds at 12 and 25$\micron$
(Boulanger \& P\'erault 1988) is believed to be thermal emission from
grains which are small enough that absorption of a single photon can
raise the grain temperature to $\sim 150\K$ for the $25\micron$ emission,
and $\sim300\K$ for the $12\micron$ emission.
Such grains contain $\sim10^2-10^3$ atoms, and must be sufficiently
numerous to account for $\sim$20\% of the total rate of absorption of
energy from starlight.\footnote{
	Assuming a Debye temperature $\Theta=420\K$, a 6 eV photon can
	heat a particle with N=277 atoms to $T=200\K$.
	}
A substantial fraction of these very small grains may be hydrocarbon
molecules, as indicated by emission bands
at 6.2, 7.7, 8.6, and 11.3$\micron$ observed from diffuse clouds by
{\it IRTS} (Onaka et al.\ 1996) and {\it ISO} (Mattila et al.\ 1996).

If the grains are primarily carbonaceous, they must contain $\sim 3-10\%$
of the carbon in the interstellar medium 
[in the model of D\'esert et al.\ (1990), 
polycyclic aromatic hydrocarbon molecules contain $\sim$9\% of the carbon]. 
The size distribution is uncertain. 
As in DL98, for our standard model ``A'' 
we assume a grain size distribution consisting of a
power law $dn/da\propto a^{-3.5}$
(Mathis, Rumpl, \& Nordsieck 1977; Draine \& Lee 1984) 
plus a log-normal distribution containing
5\% of the carbon;
50\% of the mass 
in the log-normal distribution (i.e., 2.5\% of the C atoms) 
is in grains with $N < 160$ atoms.
The grains are assumed to be disk-shaped for $N<120$, and spherical
for $N>120$.

The microwave emission from a spinning grain depends on the component
of the electric
dipole moment perpendicular to the angular velocity.
Following DL98, we assume that a neutral grain has a dipole moment
$\mu = N^{1/2}\beta$, where $N$ is the number of atoms in the grain.
Recognizing that there will be a range of dipole moments for grains of
a given $N$, we assume that 25\% of the grains have
$\beta=0.5\beta_0$, 50\% have $\beta=\beta_0$, and 25\% have $\beta=2\beta_0$.
For our standard models we will take $\beta_0=0.4\debye$, but since $\beta$
is uncertain we will examine the effects of varying $\beta_0$.

If the grain has a charge $Ze$, then it acquires an additional 
electric dipole component arising from the fact that the centroid of
the grain charge will in general be displaced from the center of mass
(Purcell 1979); we follow DL98 in assuming a characteristic displacement
of $0.1a_x$, where $a_x$ is the rms distance of the grain surface from
the center of mass, and in computing the grain charge distribution $f(Z)$
including collisional charging and photoelectric emission.

Finally we assume that the dipole moment is, on average, uncorrelated
with the spin axis, so that the mean-square dipole moment transverse
to the spin axis is
\beq
\label{eq:mueff}
\mu_\perp^2= (2/3)
\left( \beta^2 N + 0.01 Z^2e^2a_x^2 \right) ~~~.
\eeq

We will show results for 6 different grain models.
Models B and C differ from our preferred model A by having $\beta_0$
decreased or increased by a factor 2.
Model D differs by having twice as many small grains in the log-normal
component.
In model E even the smallest grains are spherical, whereas in model F
the smallest grains are assumed to be linear.

\section{Rotation of Interstellar Grains \label{sec:rotation}}

Rotation of small interstellar grains has been discussed previously,
including the
fact that the rate of rotation of very small grains would be limited by
electric dipole emission (Draine \& Salpeter 1979).
Ferrara \& Dettmar (1994) recently estimated the rotational emission from
very small grains, noting that it could be observable up to $\sim$100 GHz,
but they did not self-consistently evaluate the emission spectrum including
the effects of radiative damping.
A number of distinct physical processes act to excite and damp rotation of
interstellar grains, including collisions with neutral atoms, collisions with
ions and long-range interactions with passing ions (Anderson \& Watson 1993),
rotational emission of electric dipole radiation,
emission of infrared radiation, formation of $\HH$ molecules on the
grain surface, and photoelectric emission.
We use the rates for these processes summarized in
DL98; the results reported
here assume that $\HH$ does {\it not} form on small grains, so that
there is no contribution by $\HH$ formation to the rotation.
As discussed in DL98, observed rates of $\HH$ formation on grains limit
the $\HH$ formation torques so that they can make at most a minor contribution
to the rotation of the very small grains of interest here.

Emission in the 10-100 GHz region is dominated by grains containing
$N\ltsim10^3$ atoms.
For these very small grains under CNM conditions
(see Table \protect{\ref{tab:phases}}), rotational excitation is dominated by
direct collisions with ions and ``plasma drag''.
The very smallest grains ($N\ltsim 150$) have their rotation damped
primarily by electric dipole emission; for $150\ltsim N\ltsim 10^3$ plasma
drag dominates.

\section{Predicted Emission\label{sec:prediction}}

Following DL98, we solve for 
the rms rotation rate $\langle \nu^2\rangle^{1/2}$, which depends on
the local environmental conditions (density $n_\H$, gas temperature
$T$, fractional ionization $n_e/n_\H$, and the intensity of the
starlight background), the
grain size, and the assumed value of $\beta$ in eq.(\ref{eq:mueff}).
Then, assuming a Maxwellian distribution of angular velocities for each
grain size and $\beta$, 
we integrate over the size distribution to obtain the
emission spectrum for the grain population.

Interstellar gas is found in a number of characteristic
physical states (see, e.g., McKee 1990).
By mass, most of the gas and dust is found in the
``Warm Ionized Medium'' (WIM),
``Warm Neutral Medium'' (WNM),
``Cold Neutral Medium'' (CNM),
or in molecular clouds (MC).
In the diffuse regions (WIM,WNM,CNM) the bulk of the 
dust is heated by the general
starlight background to temperatures $\sim18\K$ where it radiates
strongly at $\sim100\micron$ -- thus the $100\micron$ emission traces
the mass of the WIM, WNM, and CNM material.
Relatively little molecular material is present at $|b|>30\deg$ where
sensitive CBR observations are directed.
Hence we consider only the WIM, WNM, and CNM phases in the present
discussion.

\begin{figure}
\epsscale{0.90}
\plotone{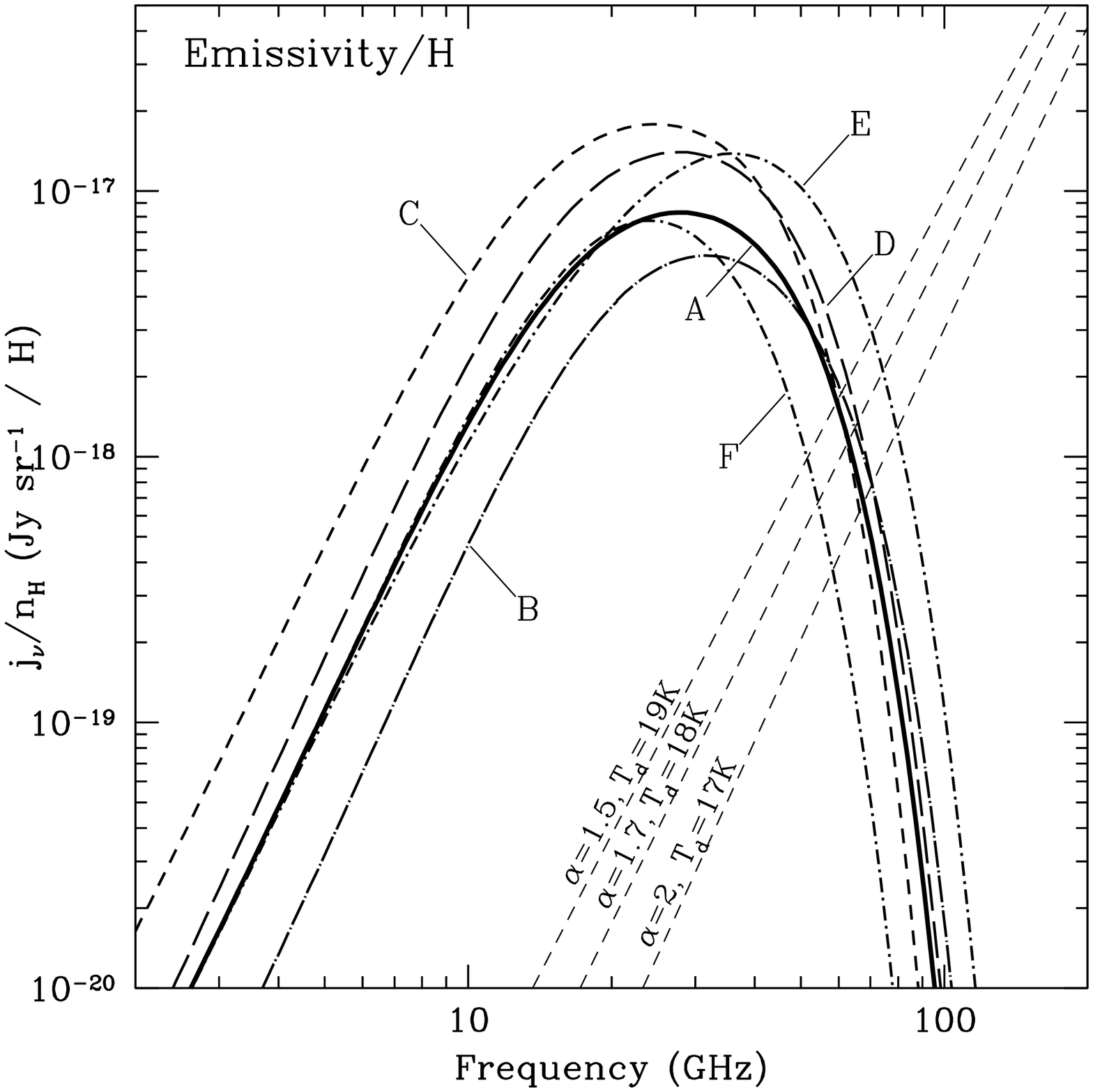}
\figcaption[f1.eps]{
	\label{fig:specs}
	Predicted rotational emissivity/H of interstellar dust
	at $|b|\gtsim 20\arcdeg$.
	We assume the mix of cloud properties given in Table 
	\protect{\ref{tab:phases}}.
	Curves are labelled by grain model, A-F
	(see Table \protect{\ref{tab:models}}).
	A is the preferred grain model.
	Estimates for the vibrational (``thermal'') emission from
	$a\gtsim 0.01\micron$ grains
	are labelled by $\alpha$ and $T_d$
	(see text).
	}
\end{figure}

In Table 1 we list the assumed properties for each of the interstellar
medium components which we consider here.
For $b>30\arcdeg$, $21\cm$ observations (Heiles 1976) indicate
$N_\H({\rm WNM})+N_\H({\rm CNM})\approx 3.4\times10^{20}(\csc b-0.2)\cm^{-2}$,
divided approximately equally between WNM and CNM phases 
(Dickey, Salpeter, \& Terzian 1978).
Dispersion measures toward pulsars in 4 globular clusters 
(Reynolds 1991) indicate
an ionized component
$N_\H({\rm WIM})\approx 5\times10^{19}\csc b\cm^{-2}$, after attributing $\sim$20\%
of the electrons to hot ionized gas (McKee 1990).
In Figure \ref{fig:specs} we show the predicted rotational 
emissivity per H nucleon, 
where we
have taken the weighted average of emission from CNM, WNM, and WIM,
according to the mass fractions in
Table \ref{tab:phases}.
The heavy curve is for our preferred grain model A; the results shown
for grain models B-F serve to illustrate 
the uncertainty in the predicted emission.
All models have the grain emissivity peaking at about $\sim$30 GHz; when
the smallest grains are taken to be spheres (model E), 
the emission peak shifts to
$\sim$35 GHz because of the reduced moment of inertia of these smallest
grains.

In addition to the rotational emission from the very small grains,
we expect continuum emission from the vibrational modes of the
larger dust grains, which are thermally excited according to the
grain vibrational temperature $T_d$.
The vibrational emission per H atom is assumed to vary as $j_\nu/n_\H = A \nu^\alpha
B_\nu(T_d)$.
We show the emissivity for $\alpha=1.5$, 1.7, and 2;
in each case
we adjust $A$ and $T_d$ so that 
$\nu j_\nu$ peaks at $\lambda=140\micron$,
with a peak emissivity 
$4\pi\nu j_\nu/n_\H = 3\times 10^{-24}\erg\s^{-1}\H^{-1}$
(Wright et al.\ 1991; cf. Fig. 5 of Draine 1995).
The resulting values of $\alpha$ and $T_d$ are within the range found
by Reach et al.\ (1995).
We will use $\alpha=1.7$ to estimate the thermal emissivity; by comparison
with the estimates for $\alpha=1.5$ and $\alpha=2$ we see that the
thermal emission at $\sim$100 GHz is uncertain by at least a factor $\sim2$.
We expect the thermal emission to 
dominate for $\nu\gtsim70$ GHz,
with the rotational emission dominant at lower frequencies.

\section{Comparison With Observations\label{sec:observations}}

\begin{figure}
\epsscale{0.90}
\plotone{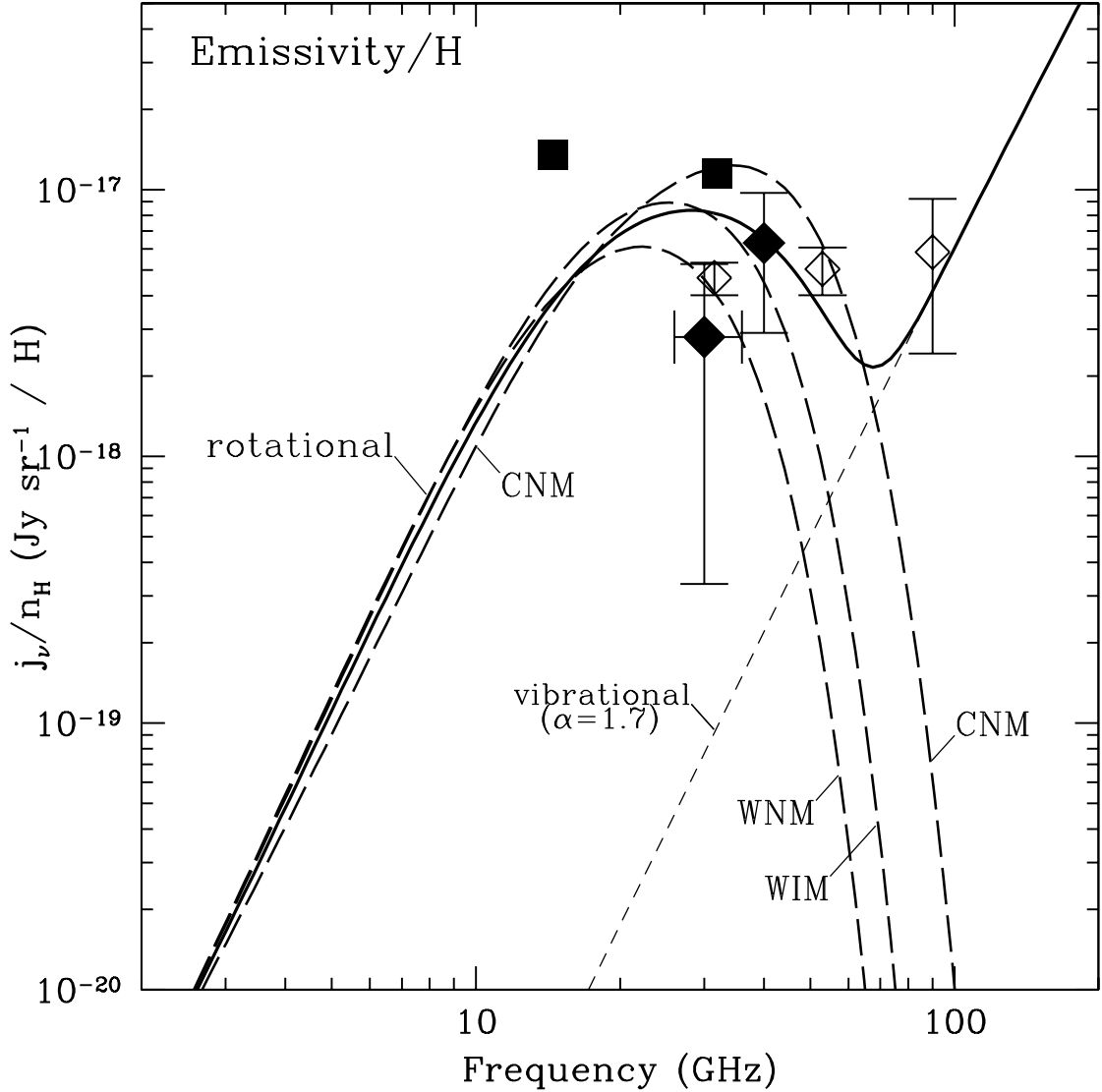}
\figcaption[f2.eps]{
	\label{fig:speca}
	Solid line is the 
	predicted emissivity/H of the interstellar medium at
	$|b|\gtsim20\arcdeg$ for the preferred grain model (``A''),
	averaged over CNM, WNM, and WIM phases of interstellar gas.
	The vibrational and rotational components of this emission
	are shown as broken lines.
	Also shown are the emissivity per H atom 
	for each of these phases.
	Symbols show observational results from
	{\it COBE} DMR (open diamonds; Kogut et al. 1996);
	Saskatoon (filled diamonds; de Oliveira-Costa et al.
	1997); and Owens Valley (filled squares; Leitch et al. 1997).
	}
\end{figure}

Figure \ref{fig:speca} shows the emission per H nucleon 
for our preferred parameters
(model A, $\alpha=1.7$), for each phase as well as the weighted average
over the three phases.
Also shown are the observational results of
Kogut et al.\ (1996) (based on cross-correlation of {\it COBE} DMR 31.5, 53, and 90 GHz
maps with {\it COBE} DIRBE 100$\micron$ maps);
de Oliveira-Costa et al.\ (1997) (based on cross-correlation of
Saskatoon 30 and 40 GHz maps with {\it COBE} DIRBE 100$\micron$ maps;
and Leitch et al.\ (1997) (based on cross-correlation of
Owens Valley mapping at 14.5 and 32 GHz
with {\it IRAS} 100$\micron$ maps.
All three papers reported strong positive correlations of
microwave emission with 100$\micron$ thermal emission from dust.
The observed correlation of 100$\micron$ emission with
21cm emission,
$I_\nu(100\micron)\approx 0.85{\rm MJy}\sr^{-1} (N_\H/10^{20}\cm^{-2})$
(Boulanger and P\'erault 1988), allows us to infer the excess
microwave emission per H atom, as shown in Fig.\ \ref{fig:speca}.

While Kogut
et al.\ attribute only $\sim$50\% of the 90 GHz signal to thermal emission,
we estimate that the 90 GHz signal is predominantly vibrational emission
from dust.  The rotational emission which we predict appears to be in
good agreement with the 30-50 GHz measurements by Kogut et al.,
de Oliveira-Costa et al., and Leitch et al.
We conclude that rotational emission from small dust grains accounts
for a substantial fraction of the ``anomalous'' Galactic emission at
30-50 GHz.

Leitch et al.\ also report excess emission at 14.5 GHz which is correlated
with 100$\micron$ emission from dust.
For our model A, rotational emission from dust grains
accounts for only $\sim$30\% of the reported excess
emission at 14.5 GHz.
Additional emission at 14.5 GHz could be obtained by
changes in our adopted parameters: model B, in which dipole moments
are increased by a factor of two, 
would be in good agreement with the 14.5 GHz result of Leitch et al.
However, the assumed dipole moment is larger than we consider likely, so
we do not favor this model.
We could of course also 
improve agreement by increasing the number of small grains in the
size range ($N\approx 100$) primarily radiating near 14.5 GHz.

We note that there may be systematic
variations in the small grain population from one region to another.
The signal reported by Leitch et al.\ at 30 GHz is about a factor of
$\sim$3 larger than the results of Kogut et al.\ and de Oliveira-Costa et al.,
both of which average over much larger areas than the 
observations of Leitch et al.
Additional observations to measure more precisely the Galactic emission
will be of great value.
We also note that
although Leitch et al.\ found no correlation of 14.5 GHz excess with
synchrotron maps at 408 MHz, synchrotron emission and rotational emission from
dust are expected to contribute approximately equally to the diffuse
background near 14 GHz (see Fig. \ref{fig:back}).
We conjecture that some of the 100$\micron$-correlated 14.5 GHz radiation
observed by Leitch et al.\ may be synchrotron emission which is enhanced
by increased magnetic field strengths near concentrations of gas 
and dust.  

\section{Why Not Free-Free? \label{sec:freefree}}

Based on the observed frequency-dependence of the excess radiation,
Kogut et al.\ proposed that it was a combination of thermal
emission from dust plus free-free emission from ionized hydrogen.
Leitch et al.\ reached the same conclusion based on their
14.5 and 32 GHz measurements.

Leitch et al.\ noted that if the proposed spatially-varying 
free-free emission originated
in gas with $T\ltsim10^4\K$, it would be accompanied by H$\alpha$
emission at least 60 times stronger than the observed variations
in the H$\alpha$ sky brightness 
(Gaustad et al.\ 1996) on the same angular scales.
Noting that the ratio of H$\alpha$ to
free-free radio continuum drops as the gas temperature is
increased above $\sim3\times10^4\K$,
Leitch et al.\ proposed that the ``anomalous'' emission
originated in gas at $T\gtsim10^6\K$, 

However, this proposal appears to be untenable.
According to Leitch et al., 
the observed emission excess would require an emission measure
$EM \approx 130 T_6^{0.4}\cm^{-6}\pc$ with 
$T_6\equiv(T/10^6\K) \gtsim 1$
near the North Celestial Pole (NCP,
$l\approx123\arcdeg$, $b\approx27.4\arcdeg$).
Kogut et al.\ and de Oliveira-Costa et al.\ reported a weaker signal at
30 GHz, corresponding to an emission measure
$\sim 40 T_6^{0.4}\cm^{-6}\pc$ toward the NCP.
The DMR observations (Kogut et al.\ 1996) cover a substantial fraction
of the sky, and would imply
an emission measure normal to
the galactic disk
$\sim 2\sin27\arcdeg\times40 T_6^{0.4}\cm^{-6}\pc$.
For $T\gtsim10^6\K$ the radiative cooling rate can be approximated
by $\sim1\times10^{-22}(T_6^{-1}+0.3T_6^{1/2})n_\H n_e \erg\cm^3\s^{-1}$
(cf. Bohringer \& Hensler 1989)
so the power radiated per disk area would
be $\sim 30(T_6^{-0.6}+0.3T_6^{0.9})\Lsol\pc^{-2}$,
far in excess of the
$0.3\Lsol\pc^{-2}$ energy input for one $10^{51}\erg$ supernova 
per 100 yr per
10 kpc radius disk.
Evidently the proposed attribution of the ``anomalous emission'' to
bremsstrahlung from
hot gas can be rejected on energetic grounds.

\section{Discussion\label{sec:discussion}}

Very small dust grains have been invoked previously to explain the
observed 3 - 60$\micron$ emission from interstellar dust
(Leger \& Puget 1984; Draine \& Anderson 1985; D\'esert, Boulanger,
\& Puget 1990).
We have calculated the spectrum of rotational emission expected from
such dust grains,
and shown that it should be detectable 
in the 10 -- 100 GHz region.
In fact, we argue that this emission has already been detected by
Kogut et al.\ (1996), de Oliveira-Costa et al.\ (1997), and Leitch et al.\ (1997).

\begin{figure}
\epsscale{0.90}
\plotone{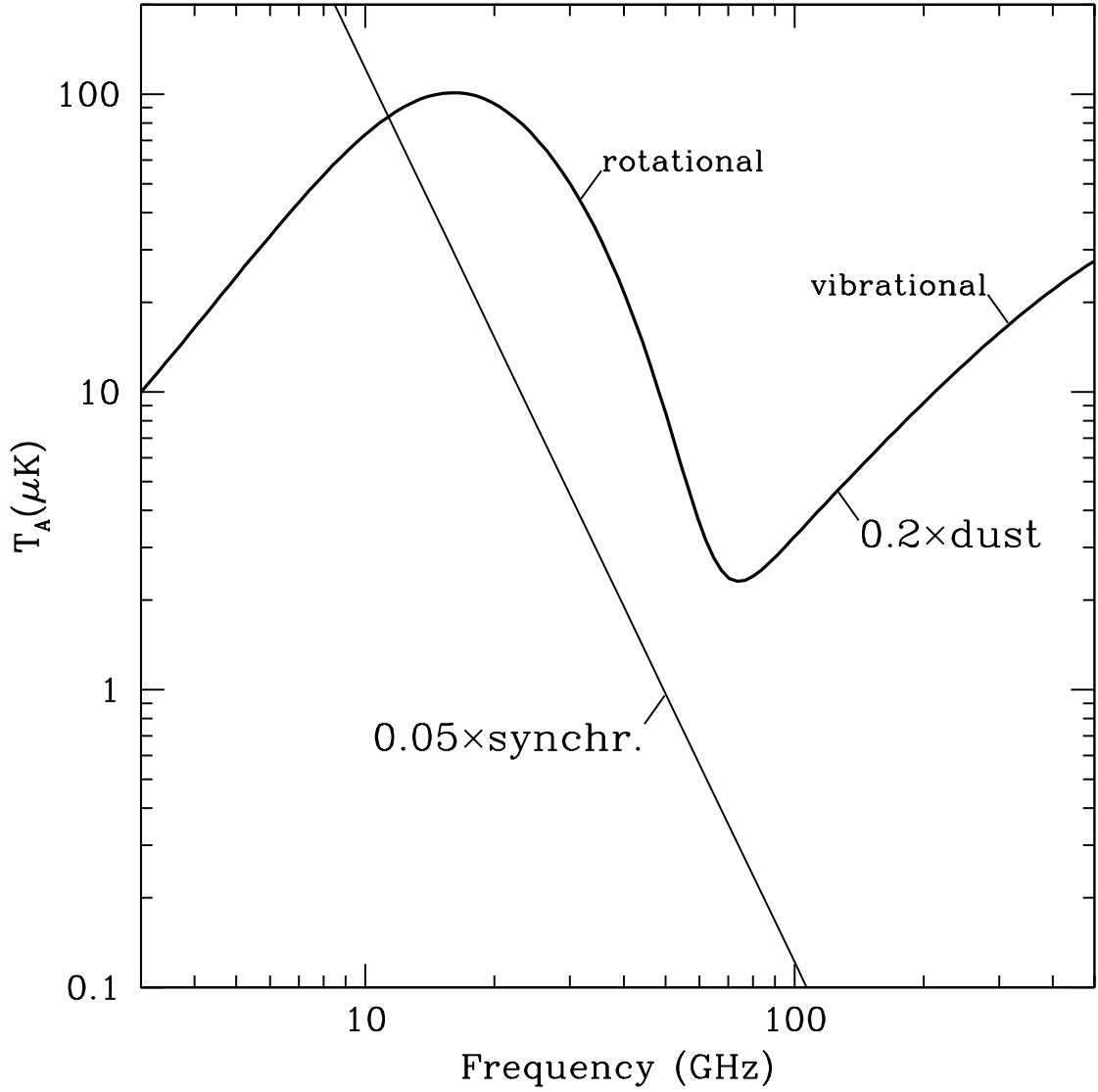}
\figcaption[f3.eps]{
	\label{fig:back}
	Estimated rms variations on scales of a few degrees 
	from galactic foregrounds
	near the North Celestial Pole (see text);
	the dust column density is assumed to vary by $\sim$20\%,
	and the synchrotron emission by $\sim$5\%.
	}
\end{figure}

Rotational emission from very small grains and ``thermal" (i.e., vibrational) 
emission by 
larger dust grains is an important foreground
which will have to be subtracted in future sensitive experiments to
measure 
angular structure in the CBR.
To illustrate the relative importance of the emission from dust, 
in Fig. \ref{fig:back} we show the estimated rms variations in Galactic
emission near the
NCP ($l=123\arcdeg$, $b=27.4\arcdeg$),
representative of intermediate galactic latitudes.
The HI column density is taken to be $N(\H^0)=6.05\times10^{20}\cm^{-2}$
(Hartmann \& Burton 1997), plus
an additional column density $N(\H^+)\sim1.3\times10^{20}\cm^{-2}$ of WIM.
We take 20\% as an estimate of the rms variations in column density
on angular scales of a few degrees.

We also show the synchrotron background near the NCP, based on a total antenna
temperature of $3.55\K$ at 1.42 GHz
(Reich 1982) and $I_\nu\propto\nu^{-1}$
at higher frequencies.
The synchrotron background is smoother than the HI; we take 5\% as
an estimate of the rms synchrotron variations on angular scales of
a few degrees.
From the standpoint of minimizing confusion with
non-CBR backgrounds, 70-100 GHz appears to be the optimal frequency window.

The ``thermal'' emission originates in larger grains, which are known to
be partially aligned with their long axes perpendicular to the local 
magnetic field; 
above 100 GHz most of the emission is thermal, and we 
estimate that this will be $\sim5\%$ linearly polarized perpendicular to the
magnetic field (Dotson 1996 has observed up to 7\% polarization at
100$\micron$ toward M17).
Below $\sim$50 GHz most of the emission is rotational.
At this time it is not clear whether the angular momentum vectors of
very small dust grains will
be aligned with the galactic magnetic field.
Ultraviolet observations of interstellar polarization 
place limits on such alignment
(Kim \& Martin 1995) but do not require it to be zero.
If the grain angular momenta tend to be aligned with the galactic
magnetic field, then the rotational emission will tend to be polarized
perpendicular to the magnetic field.
Physical processes which could produce alignment of these small grains
will be discussed in future work (Lazarian \& Draine 1998).
Even modest ($\sim$1\%) polarization of the dust emission could
interfere with efforts to measure the small polarization of the
CBR introduced by cosmological density fluctuations.

According to our model, the $\sim30\,$GHz emission and diffuse $12\micron$
emission should be correlated, as both originate in grains containing
$\sim10^2$ atoms; future satellite observations of the diffuse
background can test this.
Future observations by both ground-based experiments and satellites
such as {\it MAP} and {\it PLANCK} will
be able to characterize more precisely the intensity and spectrum of
emission from interstellar dust grains in the 10-200 GHz region.
Measurements of the 10-100 GHz emission
will constrain the abundance of very small grains, including
spatial variations, as well as their possible alignment.

\acknowledgements

We thank Tom Herbig, Ang\'elica de Oliveira-Costa, 
Lyman Page, Alex Refregier, David Spergel, 
Max Tegmark, and
David Wilkinson for helpful discussions.
We thank
Robert Lupton for the availability of the SM package.
B.T.D. acknowledges the support
of NSF grants AST-9319283 and AST-9619429, and
A.L. the support of NASA grant NAG5-2858.


\newpage

\newpage
\begin{table}
\begin{center}
\begin{tabular}{| c | c| c | c | c | c | c | }
\hline
quantity				&A	&B	&C	&D	&E	&F\\
\hline
carbon fraction\tablenotemark{a}	&0.05	&0.05	&0.05	&0.1	&0.05	&0.05	\\
$35<N<120$ shape\tablenotemark{b}	&disk	&disk	&disk	&disk	&sphere	&disk	\\
$25<N<35$ shape\tablenotemark{c}	&disk	&disk	&disk	&disk	&sphere	&linear	\\
$\beta_0$ (debye)\tablenotemark{d}&0.4	&0.2	&0.8	&0.4	&0.4	&0.4\\
\hline
\end{tabular}
\end{center}
\caption{\label{tab:models}
	Models for Very Small Grains
	}
\tablenotetext{a}{Fraction of cosmic carbon abundance in log-normal
	component (see text).}
\tablenotetext{b}{Shape of grains containing $35<N<120$ atoms (see text).}
\tablenotetext{c}{Shape of grains containing $25<N<35$ atoms (see text).}
\tablenotetext{d}{(25,50,25)\% of grains are taken to have $\beta=(0.5,1,2)\beta_0$ 
	(see eq.\ref{eq:mueff}).}
\end{table}

\begin{table}
\begin{center}
\begin{tabular}{ | c| c | c | c | }
\hline
quantity				&CNM	&WNM	&WIM	\\
\hline
mass fraction\tablenotemark{a}		&0.43	&0.43	&0.14	\\
$n_\H (\cm^{-3})$			&30	&0.4	&0.1	\\
$T (\K)$				&100.	&6000.	&8000.	\\
$\chi$					&1.	&1.	&1.	\\
$x_\H\equiv n(\H^+)/n_\H$ 		&0.0012	&0.1	&0.99	\\
$x_M\equiv n(M^+)/n_\H$\tablenotemark{b}&0.0003	&0.0003	&0.001	\\
\hline
\end{tabular}
\caption{
	Idealized phases for diffuse interstellar matter.
	\label{tab:phases}
	}
\end{center}
\tablenotetext{a}{Fraction of CNM+WNM+WIM in each phase (see text).}
\tablenotetext{b}{Relative abundance of ions other than H$^+$ or He$^+$.}
\end{table}

\pagebreak

\begin{thebibliography}{}
\bibitem[]{AW93} Anderson, N., \& Watson, W.D. 1993,
	A\&A, 270, 477
\bibitem[]{BH89} Bohringer, H., \& Hensler, G. 1989, A\&A, 215, 147
\bibitem[]{BP88} Boulanger, F., \& P\'erault, M. 1988, ApJ, 330, 964
\bibitem[]{dOC97} de Oliveira-Costa, A., Kogut, A., Devlin, M.J., 
	Netterfield, C.B., Page, L.A., \& Wollack, E.J. 1997,
	ApJ, 482, L17
\bibitem[]{DBP90} D\'esert, F.-X., Boulanger, F., \& Puget, J.L. 1990,
	A\&A, 237, 215
\bibitem[]{DST78} Dickey, J.M., Salpeter, E.E., \& Terzian, Y. 1978,
	ApJS, 36, 77
\bibitem[]{Dot19} Dotson, J.L. 1996, ApJ, 470, 566
\bibitem[]{Dr95} Draine, B.T. 1995, in
	``Physics of the Interstellar Medium and Intergalactic Medium'',
	ed. A. Ferrara, C.F. McKee, C. Heiles, and P.R. Shapiro,
	ASP Conf. Ser. 80, 133
\bibitem[]{DA85} Draine, B.T., \& Anderson, N. 1985, ApJ, 292, 494
\bibitem[]{DrLa98} Draine, B.T., \& Lazarian, A. 1998, in preparation
\bibitem[]{DL84} Draine, B.T., \& Lee, H.M. 1984, ApJ, 285, 89
\bibitem[]{DS79} Draine, B.T., \& Salpeter, E.E. 1979, ApJ, 231, 77
\bibitem[]{FD94} Ferrara, A., \& Dettmar, R.-J. 1994, ApJ, 427, 155
\bibitem[]{GMVB96} Gaustad, J.E., McCullough, P.R., \& Van Buren, D. 1996,
	PASP, 108, 823
\bibitem[]{HB97} Hartmann, D., \& Burton, W.B. 1997,
	"Atlas of Galactic Neutral Hydrogen",
        Cambridge University Press
\bibitem[]{He76} Heiles, C. 1976, ApJ, 204, 379
\bibitem[]{KM95} Kim, S.H., \& Martin, P.G. 1995, ApJ, 444, 293
\bibitem[]{Koea96} Kogut, A., Banday, A.J., Bennett, C.L.,
	Gorski, K.M., Hinshaw, G., \& Reach, W.T. 1996,
	ApJ, 460, 1
\bibitem[]{LD98} Lazarian, A., \& Draine, B.T. 1998, in preparation
\bibitem[]{LP84} Leger, A., \& Puget, J.L. 1984, A\&A, 137, L5
\bibitem[]{LRPM97} Leitch, E.M., Readhead, A.C.S., Pearson, T.J., \& 
	Myers, S.T. 1997, ApJ, 486, L23
\bibitem[]{MRN77} Mathis, J.S., Rumpl, W., \& Nordsieck, K.H. 1977,
	ApJ, 217, 425
\bibitem[]{Mat96} Mattila, K., Lemke, D., Haikala, L.K.,
	Laureijs, R.J., Leger, A., Lehtinen, K., Leinert, C.,
	Mezger, P.G. 1996, A\&A, 315, L353
\bibitem[]{McK90} McKee, C.F. 1990, in
	``The Evolution of the Interstellar Medium'',
	ed. L. Blitz,
	ASP Conf. Ser. 12, 3
\bibitem[]{OYTR96} Onaka, T., Yamamura, I., Tanab\'e, T., Roellig, T.,
	\& Yuen, L. 1996, PASJ, 48, L59
\bibitem[]{Pur79} Purcell, E.M. 1979, ApJ, 231, 404.
\bibitem[]{RDF95} Reach, W.T., et al.\ 1995, ApJ, 451, 188
\bibitem[]{Rei82} Reich, W. 1982, A\&A Suppl., 48, 219
\bibitem[]{Rey91} Reynolds, R.J. 1991, ApJ, 372, L17
\bibitem[]{Wri91} Wright, E.L., et al.\ 1991, ApJ, 381, 200
\end{thebibliography}
\end{document}